\documentclass{eptcs}
\providecommand{\event}{FOCLASA 2011}

\usepackage{verbatim}
\usepackage{amsmath}
\usepackage{subfigure}
\usepackage{graphicx}
\usepackage{colortbl}
\usepackage{multirow}
\usepackage{float}

\floatstyle{boxed}
\restylefloat{figure}
\usepackage[section]{placeins}

\usepackage{listings}

\lstdefinelanguage{rebeca}{
  morekeywords={reactiveclass, knownrebecs, statevars, main, msgsrv, main, define, LTL, CTL, boolean, int, shortint, byte, if, else, while, for, msg, reset, set, self, false, true, now, after, delay, deadline, time, env},
  otherkeywords={=>,<-,<\%,<:,>:,\#,@},
  sensitive=true,
  morecomment=[l]{//},
  morecomment=[n]{/*}{*/},
  morestring=[b]",
  morestring=[b]',
  morestring=[b]"""
}

\lstdefinestyle{scrsize}{basicstyle={\scriptsize\ttfamily}}
\lstdefinestyle{ftnsize}{basicstyle={\footnotesize\ttfamily}}

\usepackage{latexsym, stmaryrd, pxfonts}

\newcommand{\axiomrule}[1]{\raisebox{.0ex}{\scriptsize{$#1$}}}
                     
\newcommand{\sosrule}[2]{\frac{\raisebox{.7ex}{\scriptsize{$#1$}}}
                        {\raisebox{-1.0ex}{\scriptsize{$#2$}}}}

\newcommand{\trans}[1]{\,{\stackrel{{#1}}{\rightarrow}}\,}

\def\lparal{\mathbin{\setbox0=\hbox{$\|$}%
        \dimen0=\dp0 \advance\dimen0 -1.5pt \dp0=\dimen0%
        \underline{\kern-1.5pt\box0\kern1.5pt}}}

\newcommand{\leaveout}[1]{}

\def\clap#1#2{
  {
   \setbox0=\hbox{\mathify{#1}}
   \setbox1=\hbox{\mathify{#2}}
   \ifdim\wd0>\wd1
      \setbox2=\box0
      \setbox0=\box1
      \setbox1=\box2
      \fi
   \dimen0=\wd1
   \advance\dimen0 by -\wd0
   \divide\dimen0 by 2
   \dimen1=-\wd0
   \advance\dimen1 by -\dimen0
   \hskip\dimen0\box0\hskip \dimen1
   \box1
    }
  }

\newcommand{\mathdef}[1]{\relax\ifmmode #1\else $#1$\fi}

\def\mathy[[#1]]{\mathify{#1}}

\newcommand{\mathify}[1]{\ifmmode{#1}\else\mbox{$#1$}\fi}

\usepackage{latexsym, stmaryrd}

\newcommand{\rowh}[1]{$\begin{array}{c} \textbf{#1} \end{array}$}
\newcommand{\rowH}[2]{$\begin{array}{c} \textbf{#1} \\ \textbf{#2} \end{array}$}

\title{Modelling and Simulation of Asynchronous Real-Time Systems using Timed
Rebeca}

\author{Luca Aceto$^{1}$ \and Matteo Cimini$^{1}$ \and
Anna Ingolfsdottir$^{1}$ \and
Arni Hermann Reynisson$^{1}$ \and
Steinar Hugi Sigurdarson$^{1}$  \and
Marjan Sirjani$^{1,2}$
\\ $^{1}$Reykjavik University, Iceland  $~~~$  $^{2}$University of Tehran, Iran
}

\def\titlerunning{Modelling and Simulation of Asynchronous Real-Time Systems using Timed Rebeca}
\def\authorrunning{L. Aceto at al.}

\begin{document}

\maketitle


\pagenumbering{arabic}

\begin{abstract}
In this paper we propose an extension of the Rebeca language that can
be used to model distributed and asynchronous systems with timing
constraints. We provide the formal semantics of the language using
Structural Operational Semantics, and show its expressiveness by means
of examples. We developed a tool for automated translation from timed Rebeca to the
Erlang language, which provides a first implementation of timed
Rebeca. We can use the tool to set the parameters of timed Rebeca
models, which represent the environment and component variables, and
use McErlang to run multiple simulations for different settings.
Timed Rebeca restricts the modeller to a pure asynchronous actor-based paradigm, where the
structure of the model represents the service oriented
architecture, while the computational model matches the network
infrastructure. Simulation is shown to be an effective analysis support, specially where model checking faces almost immediate state explosion in an asynchronous setting.

\end{abstract}

\section{Introduction}

%

This paper presents an extension of the actor-based Rebeca
language~\cite{Sirjani04FI} that can be used to model
distributed and asynchronous systems with timing constraints. This
extension of Rebeca is motivated by the ubiquitous presence of
real-time computing systems, whose behaviour depends crucially on
timing as well as functional requirements.

A well-established paradigm for modelling the functional behaviour of
distributed and asynchronous systems is the actor model.  This
model was originally introduced by Hewitt~\cite{Hewitt72} as an
agent-based language, and is a mathematical model of concurrent
computation that treats \textit{actors} as the universal primitives of
concurrent computation \cite{Agha86}. In response to a
message that it receives, an actor can make local decisions, create
more actors, send more messages, and determine how to respond to the
next message it receives. Actors have encapsulated states and
behaviour, and are capable of  redirecting
communication links through the exchange of actor identities.
Different interpretations, dialects and extensions of actor models
have been proposed in several domains and are claimed to be the most
suitable model of computation for the dominating applications,
such as multi-core programming and web services
\cite{Hewitt07-commitment}.

$\underline{Re}active~~O\underline{b}j\underline{ec}ts~~ L\underline{a}nguage$,
$Rebeca$~\cite{Sirjani04FI}, is an operational interpretation of the
actor model with formal semantics and model-checking tools.
Rebeca is designed to bridge the gap between formal methods and
software engineers. The formal semantics of Rebeca is a solid basis
for its formal verification. Compositional and modular verification,
abstraction, symmetry and partial-order reduction have been
investigated for verifying Rebeca models. The theory underlying these
verification methods is already established and is embodied in
verification
tools~\cite{jaghoor-sirjani-mousavi-movaghar-09-Acta,Sirjani05JUCS,Sirjani04FI}.
With its simple, message-driven and object-based computational model,
Java-like syntax, and a set of verification tools, Rebeca is an
interesting and easy-to-learn model for practitioners.

\paragraph{Motivation and Contribution.}
Although actors are attracting more and more attention both in
academia and industry, little has been done on timed actors and even
less on analyzing timed actor-based models.
In this work we present
\begin{itemize}
\item  timed Rebeca by extending Rebeca with time constraints,
\item the formal semantics of timed Rebeca using Structural Operational Semantics (SOS)~\cite{Plotkin81},
\item a tool for mapping timed Rebeca models to Erlang, and
\item experimental results from the simulation of timed Rebeca models using  McErlang \cite{McErlang}.
\end{itemize}

The contribution of this work is offering a pure asynchronous actor-based modelling language with timing primitives and analysis support.
Timed Rebeca can be used in a model-driven methodology in which the
designer builds an abstract model where each component is a reactive
object communicating through non-blocking asynchronous messages. The
structure of the model can very well represent service oriented
architectures, while the computational model matches the network
infrastructure. Hence the model captures faithfully the behaviour
of the system in a distributed and asynchronous world.

\paragraph{Comparison with other timed models.} Comparing with the well-established timed models, like timed automata \cite{Alur1994}, TCCS \cite{YiTCCS91b}, and real-time Maude \cite{OlveczkyM02}, timed Rebeca offers an actor-based syntax and a built-in actor-based computational model, which restricts the style of modelling to an event-based concurrent object-based paradigm.
Modelling time-related features in computational models has been studied for a long time \cite{Baker78actorsystems,Alur1994}; while we have no claims of improving the expressiveness of timed models, we believe that our model is highly usable due to its actor-based nature and Java-like syntax.
The usability is due to the one to one correspondence between the entities of the real world and the objects in the model, and the events and actions of the real world and the computational model. Moreover, the  syntax of the language is familiar for  software engineers and practitioners.

\paragraph{Comparison with other timed actor models.}
We know of a few other timed actor-based modelling languages \cite{RenA95,Nielsen96,DBLP:journals/corr/abs-1009-4262} that we will explain in more detail in the related work section. In  \cite{RenA95} a central synchronizer acts like a coordinator and  enforces the real-time and synchronization constraints (called interaction constraints). The language for the coordinated actors is briefly proposed in \cite{Nielsen96}; however, the main focus is having reusable real-time actors without hardwired interaction constraints. The constraints declared within the central synchronizer in this line of work can be seen as the required global properties of a timed Rebeca model. We capture the architecture and configuration of a system via a timed Rebeca model and then we can check whether the global constraints are satisfied. The language primitives that we use to extend Rebeca are consistent with the proposal in \cite{Nielsen96}.
The primitives proposed in \cite{DBLP:journals/corr/abs-1009-4262} are different from ours; they introduced an \emph{await} primitive where we keep the asynchronous nature of the model.

\paragraph{Analysis support.}
In order to analyze timed Rebeca models, we developed a tool to facilitate their simulation. In a parallel project \cite{javadMSThesis2010}, a mapping from timed Rebeca to timed automata is developed and UPPAAL \cite{UppaalSite} is used for model checking. The asynchronous nature of Rebeca models causes state explosion while model checking even for small models. One solution is using a modular approach like in \cite{Jaghoori07nwpt}. Here, we selected an alternative solution as a complementary tool for analysis. Using our tool we can translate
a timed Rebeca model to Erlang \cite{ErlangSite}, set the parameters
which represent the environment and component variables, and run
McErlang \cite{McErlang} to simulate the model.
The tool allows us to change the settings of different timing parameters and rerun the
simulation in order to investigate different scenarios, find
potential bugs and problems, and optimize the model by manipulating
the settings. The parameters can be timing
constraints on the local computations (e.g., deadlines for
accomplishing a requested service), computation time for providing a
service, and frequency of a periodic event. Parameters can also
represent network configurations and delays.
In our experiments we could find timing problems that caused missing a deadline, or an unstable state in the system.

The formal semantics presented in this paper is the basis for the correct mapping from timed Rebeca to Erlang.
The detailed mapping, and the tool together with some examples can be found at \cite{IceRoseSite}.

Our choice to use the actor-based programming language Erlang is also based on the idea of covering the whole life cycle of the system in future, and of providing a refinement step for implementing the code from our timed Rebeca model.




\section{Related Work}
Different approaches are used in designing formal modelling
languages for real-time systems.
The model of timed automata, introduced by Alur and Dill \cite{Alur1994}, has established itself as a classic formalism for modelling real-time systems.
The theory of timed automata is a timed extension of automata theory, using clock constraints on
both locations and transitions.
%
In many other cases the proposed modelling languages for real-time
systems are extensions of existing languages with real-time
concepts---see, for example, TCCS \cite{YiTCCS91b} and Real-time Maude
\cite{OlveczkyM02}.

 A real-time actor model, RT-synchronizer, is proposed in
\cite{RenA95}, where a centralized synchronizer is responsible for
enforcing real-time relations between events.  Actors are extended
with timing assumptions, and the functional behaviours of actors and
the timing constraints on patterns of actor invocation are
separated. The semantics for the timed actor-based language is given
in \cite{Nielsen96}. Two positive real-valued constants, called
\emph{release time} and \emph{deadline}, are added to the \emph{send}
statement and are considered as the earliest and latest time when the
message can be invoked relative to the time that the method executing the send is invoked. In Timed Rebeca, we have the constructs
\emph{after} and \emph{deadline}, which are representing the same
concepts, respectively, except that they are relative to the time that the message (itself) is sent.
So, it more directly reflects the computation architecture including the network delays.
 In our language, it is also possible to
consider a time \emph{delay} in the execution of a computation where in \cite{Nielsen96} it is possible to specify an upper bound on the execution time of a method.  While
RT-synchronizer is an abstraction mechanism for the declarative
specification of timing constraints over groups of actors, our model
allows us to work at a lower level of abstraction. Using timed Rebeca,
a modeller can easily capture the functional features of a
system, together with the timing constraints for both computation and
network latencies, and analyze the model from various points of view.

There is also some work on schedulability analysis of actors
\cite{NigroP01}, but this is not applied on a real-time actor
language. Time constraints are considered separately.
Recently, there have been some studies on schedulability analysis for Rebeca models
\cite{Jaghoori08jlap}. This work is
based on mapping Rebeca models to timed automata and using UPPAAL to
check the schedulability of the resulting models. Deadlines are
defined for accomplishing a service and each task spends a certain
amount of time for execution. In the above-mentioned papers, modelling
of time is not incorporated in the Rebeca language.

Creol is a concurrent object-oriented language with an operational
semantics written in an actor-based style, and supported by a language
interpreter in the Maude system.
In \cite{DBLP:conf/fsen/BoerCJ09}, Creol is extended by adding
best-case and worst-case execution time for each statement, and a
deadline for each method call. In addition, an object is assigned a
scheduling strategy to resolve the nondeterminism in selecting from
the enabled processes. This work is along the same lines as the one
presented in~\cite{Jaghoori08jlap} and the focus is on schedulability
analysis, which is carried out in a modular way in two steps: first
one models an individual object and its behavioural interface as timed
automata, and then one uses UPPAAL to check the schedulability
considering the specified execution times and the deadlines. In this
work, network delays are not considered, and the execution time is
weaved together with the statements in a fine-grained way.

In \cite{DBLP:journals/corr/abs-1009-4262} a timed version of Creol is
presented in which the only additional syntax is read-only access to
the global clock, plus adding a data-type \emph{Time} together with
its accompanying operators to the language. Timed behaviour is
modelled by manipulating the \emph{Time} variables and via the
\emph{await} statement in the language.


\section{Timed Rebeca} \label{sec:timedrebeca}

A Rebeca model consists of a set of {\em reactive classes} and the
{\em main} program in which we declare reactive objects, or rebecs, as
instances of {\em reactive classes}.  A reactive class has an argument
of type integer, which denotes the length of its message queue.  The
body of the reactive class includes the declaration for its {\em known
rebecs}, variables, and methods (also called message servers). Each
method body consists of the declaration of local variables and a
sequence of statements, which can be assignments, {\em if} statements,
rebec creation (using the keyword {\em new}), and method calls. Method
calls are sending asynchronous messages to other rebecs (or to self)
to invoke the corresponding message server (method). Message passing
is fair, and messages addressed to a rebec are stored in its message
queue. The computation takes place by taking the message from the
front of the message queue and executing the corresponding message
server~\cite{Sirjani04FI}.

\paragraph{Timing features in an asynchronous and distributed setting.}

To decide on the timing primitives to be added to the Rebeca syntax, we first considered the different timing features that a modeller might need to address in a message-based, asynchronous and distributed setting. These features (like the computation time, or periodic events) can be common in any setting.
\begin{enumerate}
\item
  \textbf{Computation time}: the time needed for a computation to take place.
\item
  \textbf{Message delivery time}: the time needed for a message to travel between two objects, that depends on the network delay (and possibly other parameters).
\item
  \textbf{Message expiration}: the time within which a message is still valid. The message can be a request or a reply to a request (a request being served).
\item
  \textbf{Periods of occurrences of events}: the time periods for periodic events.
\end{enumerate}

We introduce an extension of Rebeca with real-time primitives to be able to address the above-mentioned timing features.
In timed Rebeca model, each rebec has its own local clock, which can be considered as synchronized distributed clocks\footnote{In this paper we do not address the problem of distributed clock synchronization; several options and protocols for establishing clock synchronization in a distributed system are discussed in the literature, including \cite{Tanenbaum2002}.}.
 Methods are
still executed atomically, but we can model passing of time while
executing a method.  Instead of a message queue for each rebec, we
have a bag containing the messages that are sent.
The timing primitives that are added to the syntax of Rebeca are {\em delay},
{\em now}, {\em deadline} and {\em after}.
Figure \ref{RebecaSyntax} shows the grammar for Timed Rebeca.
The {\em delay} statement models the passing of time for a rebec
during execution of a method (computation time), and {\em now} returns the local time of
the rebec.
  The keywords {\em after} and {\em deadline} can only be
used in conjunction with a method call.
Each rebec knows about
its local time and can put {\em deadline} on the messages that are sent declaring that the message will not be valid
after the deadline (modelling the message expiration). 
 The {\em after} primitive, attached to a message, can be used to declare  a constraint on the earliest time
at which the message can be served (taken from the message bag by the receiver
rebec).  The modeller may use these constraints for various purposes, such
as modelling the network delay or modelling a periodic event.

The messages that are sent are
put in the message bag together with their time tag and {\em deadline}
tag. The scheduler decides which message is to be executed next based
on the time tags of the messages. The time tag of a message is the
value of {\em now} when the message was sent, with the value of
the argument of the {\em after} added to it when the message is
augmented with an {\em after}. The intuition is that a message
cannot be taken (served) before the time that the time tag determines.

The progress of time is modeled locally by the delay statement. Each delay
statement within a method body increases the value of the local
time (variable {\em now}) of the respective rebec by the amount of its
argument.  When we reach a {\em call} statement (sending a message), we put that
message in the message bag augmented with a time tag. The local time of a
rebec can also be increased when we take a message from the bag to execute the
corresponding method.


The scheduler takes a message from the message bag, executes the
corresponding message server atomically, and then takes another
message. Every time the scheduler takes a message for execution, it
chooses a message with the least time tag. Before the execution of the
corresponding method starts, the local time ({\em now}) of the
receiver rebec is set to the maximum value between its current time
and the time tag of the message. The current local time of each rebec
is the value of {\em now}. This value is frozen when the method
execution ends until the next method of the same rebec is taken for
execution.

\begin{figure}
\begin{align*}
Model &\Coloneqq EnvVar^* ~ Class^* ~ Main \qquad EnvVar \Coloneqq \mathbf{env} ~ T ~ \langle v \rangle ^+;\\
Main &\Coloneqq \mathbf{main} ~ \{ ~ InstanceDcl^* ~ \} \qquad InstanceDcl \Coloneqq C ~ r(\langle r \rangle ^*):(\langle c \rangle ^*);\\
Class &\Coloneqq \mathbf{reactiveclass} ~ C ~ \{ ~ KnownRebecs ~ Vars ~ MsgSrv^* ~ \}\\
KnownRebecs &\Coloneqq \mathbf{knownrebecs} ~ \{ ~ VarDcl^* ~ \} ~ Vars \Coloneqq \mathbf{statevars} ~ \{ ~ VarDcl^* ~ \} ~ VarDcl \Coloneqq T ~ \langle v \rangle ^+;\\
MsgSrv &\Coloneqq \mathbf{msgsrv} ~ M(\langle T ~ v \rangle ^*) ~ \{ ~ Stmt^* ~ \}\\
Stmt &\Coloneqq v = e; ~ | ~ r = new ~ C(\langle e \rangle ^*); ~ | ~ Call; ~ | ~ if ~ (e) ~ MSt ~ [else ~ MSt] ~ | ~ \mathbf{delay}(t); ~ | ~ \mathbf{now}();\\
Call &\Coloneqq r.M(\langle e \rangle ^*) ~ [\mathbf{after}(t)] ~ [\mathbf{deadline}(t)]\\
MSt &\Coloneqq \{ ~ Stmt^* ~ \} ~ | ~ Stmt
\end{align*}
\caption{Abstract syntax of Timed Rebeca. Angle brackets $\langle$...$\rangle$ are used as meta parenthesis, superscript $+$ for repetition more than once, superscript $*$ for repetition zero or more times, whereas using $\langle$...$\rangle$ with repetition denotes a comma separated list.  Brackets $[...]$ show being optional. Identifiers $C$, $T$, $M$, $v$, $c$, and $r$ denote class, type, method, variable, constant, and rebec names, respectively;  and $e$ denotes an (arithmetic, boolean or nondetermistic choice) expression.}
\label{RebecaSyntax}
\end{figure}

%
The arguments of {\em after} and {\em delay} are relative values, but
when the corresponding messages are put in the message bag their tags
are absolute values, which are computed by adding the relative values
of the arguments to the value of the variable {\em now} of the sender
rebec (where the messages are sent).  To summarize, Timed Rebeca
extends Rebeca with the following four constructs.

\begin{itemize}

\item {\bf Delay}: {\em delay(t)}, where $t$ is a positive natural number, will
increase the value of the local clock of the respective rebec by the amount $t$.

\item {\bf Now}: {\em now()} returns the time of the local clock of the rebec from
which it is called.

\item {\bf Deadline}: {\em r.m() deadline(t)}, where $r$ denotes a rebec name, $m$
denotes a method name of $r$ and $t$ is a  natural number, means  that
the message $m$ is sent to the rebec $r$ and is put in the message bag. After
$t$ units of time the message is not valid any more and is purged from the bag.
Deadlines are used to model message expirations (timeouts).

\item {\bf After}: {\em r.m() after(t)}, where $r$ denotes a rebec
name, $m$ denotes a method name of $r$ and $t$ is a  natural
number, means that the message $m$ is sent to the rebec $r$ and is put
in the message bag. The message cannot be taken from the bag before
$t$ time units have passed.  After statements can be used to model
network delays in delivering a message to the destination, and also periodic events.

\end{itemize}

\paragraph{Ticket Service Example} We use a ticket service as a running example
throughout the article. Listing \ref{lst:timedticketservice} shows this example
written in Timed Rebeca.  The ticket service model consists of two reactive
classes: $Agent$ and $TicketService$. Two rebecs, $ts1$ and $ts2$, are
instantiated from the reactive class $TicketService$, and one rebec $a$ is
instantiated from the reactive class $Agent$. The agent $a$ is initialized by
sending a message $findTicket$ to itself in which a message $requestTicket$ is
sent to the ticket service $ts1$ or $ts2$ based on the parameter passed to
$findTicket$.  The deadline for the message $requestTicket$ to be served is
{\em requestDeadline} time units. Then, after {\em checkIssuedPeriod} time
units the agent will check if it has received a reply to its request by sending
a $checkTicket$ message to itself, modelling a periodic event.  There is no
receive statement in Rebeca, and all the computation is modeled via
asynchronous message passing, so, we need a periodic check.  The $attemptCount$
variable helps the agent to keep track of the ticket service rebec that the
request is sent to. The $token$ variable allows the agent to keep track of
which incoming $ticketIssued$ message is a reply to a valid request.  When any
of the ticket service rebecs receives the $requestTicket$ message, it will
issue the ticket after {\em serviceTime1} or {\em serviceTime2} time units, which is
modelled by sending $ticketIssued$ to the agent with the $token$ as parameter.
The expression $?(serviceTime1,serviceTime2)$ denotes a nondeterministic choice
between $serviceTime1$ and $serviceTime2$ in the {\em assignment} statement.
Depending on the chosen value, the ticket service may or may not be on time for
its reply.

\lstset{language=rebeca,style=scrsize}
\begin{lstlisting}[language=rebeca,caption=A Timed Rebeca model of the ticket service example,label=lst:timedticketservice]
env int requestDeadline, checkIssuedPeriod, retryRequestPeriod, newRequestPeriod, serviceTime1, serviceTime2; 

reactiveclass Agent {
  knownrebecs { TicketService ts1; TicketService ts2; }
  statevars { int attemptCount; boolean ticketIssued; int token; }
  msgsrv initial() { self.findTicket(ts1); } // initialize system, check 1st ticket service
  msgsrv findTicket(TicketService ts) {
    attemptCount += 1; token += 1;
    ts.requestTicket(token) deadline(requestDeadline); // send request to the TicketService
    self.checkTicket() after(checkIssuedPeriod);      // check if the request is replied
  }
  msgsrv ticketIssued(int tok) { if (token == tok) { ticketIssued = true; } }
  msgsrv checkTicket() {
    if (!ticketIssued && attemptCount == 1) {                // no ticket from 1st service,
      self.findTicket(ts2);                                  // try the second TicketService 
    } else if (!ticketIssued && attemptCount == 2) {         // no ticket from 2nd service,
      self.retry() after(retryRequestPeriod);                // restart from the first TicketService 
    } else if (ticketIssued) {                               // the second TicketService replied, 
      ticketIssued = false;
      self.retry() after(newRequestPeriod);                  // new request by a customer
    }
  }
  msgsrv retry() {
    attemptCount = 0; self.findTicket(ts1);             // restart from the first TicketService
  }
}

reactiveclass TicketService {
  knownrebecs { Agent a; }
  msgsrv initial() { }
  msgsrv requestTicket(int token) {
    int wait = ?(serviceTime1,serviceTime2);           // the ticket service sends the reply
    delay(wait);                                       // after a non-determinstic delay of
    a.ticketIssued(token);                             // either serviceTime1 or serviceTime2
  } 
}

main {
  Agent a(ts1, ts2):();                      // instantiate agent, with two known rebecs 
  TicketService ts1(a):();                   // instantiate 1st and 2nd ticket services, with 
  TicketService ts2(a):();                   // the agent as their known rebecs                       
}
\end{lstlisting}

\subsection{Structural Operational Semantics for Timed Rebeca}

In this section we provide an SOS semantics for Timed Rebeca in the
style of Plotkin~\cite{Plotkin81}. The behaviour of Timed Rebeca
programs is described by means of the transition relation
$\rightarrow$ that describes the evolution of the system.

The states of the system are pairs $(Env, B)$, where $Env$ is a finite
set of environments and $B$ is a bag of messages. For each rebec $A$
of the program there is an environment $\sigma_A$ contained in $Env$,
that is a function that maps variables to their values. The
environment $\sigma_A$ represents the private store of the rebec
$A$. Besides the user-defined variables, environments also contain  the
value for the special variables  \emph{self}, the name of the rebec, {\em now}, the current time, and {\em
sender}, which keeps track of the rebec that invoked the method that
is currently being executed. The environment $\sigma_A$ also maps
every method name of $A$ to its body.

The bag contains an unordered collection of messages. Each message is
a tuple of the form $(A_{i}, m(\overline{v}), A_j, TT, DL)$.
Intuitively, such a tuple says that at time $TT$ the sender $A_j$ sent the message to the rebec $A_i$ asking it to execute its method $m$ with actual parameters $\overline{v}$. Moreover this message expires at time $DL$.

The system transition relation $\rightarrow$ is defined by the rule
 {\em scheduler}: 
\begin{align*}
(scheduler) & ~ \sosrule{
(\sigma_{A_{i}}(m), \sigma_{A_{i}}[\mbox{now} = \max(TT, \sigma_{A_{i}}(now)),~[\overline{arg}=\overline{v}],~sender=A_{j}],~ Env,~ B) \trans{\tau} (\sigma_{A_{i}}',~ Env', B')}
{(\{\sigma_{A_{i}}\} \cup Env, \{(A_{i}, m(\overline{v}), A_j,TT, DL)\} \cup B) \rightarrow (\{\sigma_{A_{i}}'\} \cup Env',~ B')}
& C
\end{align*}
where the condition $\mathcal{C}$ is defined as follows: $\sigma_{A_{i}}$ is not contained in $Env$, and $(A_{i}, m(\overline{v}), A_j, TT, DL) \notin B$, and $\sigma_{A_{i}}(now) \leq DL$, and $TT \leq \min(B)$.
The {\em scheduler} rule allows the system to progress by picking up
 messages from the bag and executing the corresponding methods.
The third side condition of the rule, namely $\sigma_{A_{i}}(now) \leq
 DL$, checks whether the selected message carries an expired deadline,
 in which case the condition is not satisfied and the message cannot
 be picked. The last side condition is the predicate $TT \leq min(B)$,
 which shows that the time tag $TT$ of the selected message has been the smallest
 time tag of all the messages for all the rebecs $A_i$ in the bag $B$. The
 premise executes the method $m$, as described by the transition
 relation $\trans{\tau}$, which will be defined below. The method body
 is looked up in the environment of $A_i$ and is executed in the
 environment of $A_i$ modified as follows: (1) The variable {\em sender} is set to the sender of the message. (2) In executing the method $m$, the formal parameters $\overline{arg}$ are set to the values of the actual parameters $\overline{v}$. Methods of arity $n$ are supposed to have $arg_1, arg_2, \ldots , arg_n$ as formal parameters. This is without loss of generality since such a change of variable names can be performed in a pre-processing step for any program. (3) The variable {\em now} is set to the maximum between the current time of the rebec and the time tag of the selected message.

The execution of the methods of rebec $A_i$ may change the private
store of the rebec $A_i$, the bag $B$ by adding messages to it and the
list of environments by creating new rebecs through {\em new}
statements. Once a method is executed to completion, the resulting bag
and list of environments are used to continue the progress of the
whole system.

The transition relation $\trans{\tau}$ describes the execution of
methods in the style of natural semantics~\cite{Kahn87}.  (See
Figure~\ref{fig:transitions} for selected rules. The full set of rules
may be found in Appendix~\ref{App:methodrules}.) Since in this kind of
semantics the whole computation of a method is performed in a single
step, this choice perfectly reflects the atomic execution of methods
underlying the semantics of the Rebeca language. The general form of
this type of transition is $(S, \sigma, Env, B) \trans{\tau} (\sigma',
Env', B')$.  A single step of $\trans{\tau}$ consumes all the code $S$
and provides the value resulting from its execution. Carrying the bag
$B$ is important because new messages may be added to it during the
execution of a statement $S$. Also $Env$ is required because {\em new}
statements create new rebecs and may therefore add new environments to
it. In the semantics, the local environment $\sigma$ is separated from
the environment list $Env$ for the sake of clarity. The result of the
execution of the method thus amounts to the modified private store
$\sigma'$, the new list of environments $Env'$ and the new bag $B'$.

The rules for assignment, conditional statement and sequential
composition
are
standard. The rules for the timing primitives deserve some
explanation.
\begin{itemize}
\item
Rule {\em msg} describes the effect of method invocation
statements. For the sake of brevity, we limit ourselves to presenting
the rule for method invocation statements that involve both the {\em
after} and {\em deadline} keywords. The semantics of instances of that
statement without those keywords can be handled as special cases of
that rule by setting the argument of {\em after} to zero and that of
{\em deadline} to $+\infty$, meaning that the message never expires.
Method invocation statements put a new message in the bag, taking care
of properly setting its fields. In particular the time tag for the
message is the current local time, which is the value of the variable
$now$, plus the number $d$ that is the parameter of the {\em after}
keyword.
\item Delay statements change the private variable $now$ for the
considered rebec.
\end{itemize}
Finally, the creation of new rebecs is handled by the rule {\em
create}. A fresh name $A$ is used to identify the newly created rebec
and is assigned to {\em varname}. A new environment $\sigma_{A}$ is
added to the list of environments. At creation time, $\sigma_{A}$ is
set to have its method names associated to their code. A
message is put in the bag in order to execute the {\em initial}
method of the newly created rebec.

\begin{figure}
\begin{align*}
(msg) & ~ \axiomrule{(varname.m(\overline{v}) ~ after(d) ~deadline( DL), \sigma, Env, B)} \\
& ~ \axiomrule{\trans{\tau} (\sigma, Env, \{(\sigma(varname), m(eval(\overline{v}, \sigma)), \sigma(self), \sigma(now)+d, \sigma(now)+DL)\} \cup B)} \\
(delay)& ~ \axiomrule{(delay(d), \sigma, Env, B) \trans{\tau} (\sigma[now = \sigma(now) + d], Env, B)} \\
(create)& ~ \axiomrule{(varname = new~ O(\overline{v}), \sigma, Env, B)} \\
& ~ \axiomrule{\trans{\tau} (\sigma[varname = A],  \{\sigma_{A}[now = \sigma(now), ~ \mathit{self} = A]\} \cup Env, \{(A, initial(eval(\overline{v}, \sigma)), \sigma(self),\sigma(now), +\infty)\} \cup B)}
\end{align*}
\caption{Selected Method-Execution Transition Rules. In rule {\em create}, the rebec name $A$ should not appear in the range of the environment $\sigma$. The function $eval$ evaluates expressions in a given environment in the expected way. In each rule, we assume that $\sigma$ is not contained in $Env$.}
\label{fig:transitions}
\end{figure}

\section{Mapping from Timed Rebeca to Erlang}

In this section, we present a translation from the fragment of Timed
Rebeca without rebec creation to Erlang (for an extended explanation and a more formal description see \cite{IceRoseSite}). The motivation for
translating Timed Rebeca models to Erlang code is to be able to use
McErlang \cite{McErlang} to run experiments on the models. This
translation also yields a first implementation of Timed Rebeca.

McErlang is a model-checking tool written in Erlang to verify distributed
programs written in Erlang. It supports Erlang datatypes, process
communication, fault detection and fault tolerance and the Open Telecom
Platform (OTP) library, which is used by most Erlang programs. The verification
methods range from complete state-based exploration to simulation, with
specifications written as LTL formulae or hand-coded runtime monitors. This
paper focuses on simulation  since model checking with real-time
semantics is not yet offered by McErlang.
Note, however, that our translation opens the possibility of model
checking (untimed) Rebeca models using McErlang, which is not the
subject of this paper.

\paragraph{Erlang Primer}
\lstset{language=erlang,style=ftnsize}

Erlang is a dynamically-typed general-purpose programming language,
which was designed for the implementation of distributed, real-time
and fault-tolerant applications. Originally, Erlang was mostly used
for telephony applications such as switches. Its concurrency model is
based on the actor model.

Erlang has few concurrency and timing primitives:

\begin{itemize}

\item \lstinline|Pid = spawn(Fun)| creates a new process that
evaluates the given function \lstinline|Fun| in parallel with the
process that invoked spawn.

\item \lstinline|Pid ! Msg| sends the given message \lstinline|Msg| to
the process with the identifier \lstinline|Pid|.

\item \lstinline|receive ... end| receives a message that has been
sent to a process; message discrimination is based on pattern
matching.

\item \lstinline|after| is used in conjunction with a \lstinline|receive| and is followed by a timeout block as shown in Listing
  \ref{lst:receivetimeout}, after the specified time (deadline for receiving the required pattern) the process executes the timeout block
\item
  \lstinline|erlang:now()| returns the current time of the process

\end{itemize}
When a process reaches a \lstinline|receive|
expression it looks in the queue and takes a message that matches
the pattern if the corresponding guard is true. A guard is a boolean
expression, which can include the variables of the same process.  The
process looks in the queue each time a message arrives until the
timeout occurs.

\begin{lstlisting}[float=tbp, caption=Syntax of a receive with timeout., label=lst:receivetimeout]
receive
  Pattern1 when Guard1 -> Expr1;
  Pattern2 when Guard2 -> Expr2;
  ...
after
  Time -> Expr
end
\end{lstlisting}



\paragraph{Mapping}
The abstract syntax for a fragment of Erlang that is required to
present the translation is shown in Figure \ref{ErlangSyntax}.  Table
\ref{tab:erlmapping} offers an overview of how a construct in one
language relates to one in the other. We discuss the general
principles behind our translation in more detail below.

\begin{figure}
\begin{align*}
Program &\Coloneqq Function^* ~ ~ ~ Function \Coloneqq v(Pattern^*) \rightarrow e\\
Expr &\Coloneqq e_1 ~ op_e ~ e_2 ~ | ~ e(\langle e \rangle ^*) ~ | ~ e_1 ~ \mathbf{!} ~ e_2 ~ | ~ e_1 ~ \mathbf{,} ~ e_2 ~ | ~ Pattern ~ \mathbf{=} ~ e ~| ~ \mathbf{case} ~ e ~ \mathbf{of} ~ Match ~ \mathbf{end} ~ | ~ \mathbf{receive} ~ Match ~  \mathbf{end}\\
&| ~ \mathbf{receive} ~ Match ~ \mathbf{after} ~ Time ~ \rightarrow ~ e ~ \mathbf{end}
     | ~ \mathbf{if} ~ \langle Match \rangle ^* \mathbf{end} ~  | ~ BasicValue ~ | ~ v ~ | ~ \mathbf{\{} \langle e \rangle ^* \mathbf{\}} ~ | ~ \mathbf{[} \langle e \rangle ^* \mathbf{]}\\
Match &\Coloneqq Pattern ~ \mathbf{when} ~ Guard \rightarrow e \\
Pattern &\Coloneqq v ~ | ~ BasicValue ~ | ~ \mathbf{\{} \langle Pattern \rangle ^* \mathbf{\}} ~ | ~ \mathbf{[} \langle Pattern \rangle ^* \mathbf{]} ~ ~ ~ Time \Coloneqq \mathrm{int} \\
Value &\Coloneqq BasicValue ~ | ~ \mathbf{\{} \langle Value \rangle ^* \mathbf{\}} ~ | ~ \mathbf{[} \langle Value \rangle ^* \mathbf{]} ~ ~ ~ BasicValue \Coloneqq \mathrm{atom} ~ | ~ \mathrm{number} ~ | ~ \mathrm{pid} ~ | ~ \mathrm{fid}\\
Guard &\Coloneqq g_1 ~ op_g ~ g_2 ~ | ~ BasicValue ~ | ~ v ~ | ~ g( \langle g \rangle ^*) ~ | ~ \mathbf{\{} \langle g \rangle ^* \mathbf{\}} ~ | ~ \mathbf{[} \langle g \rangle ^* \mathbf{]}
\end{align*}
\caption{Abstract syntax of a relevant subset of Erlang. Angle
brackets $\langle$...$\rangle$ are used as meta parenthesis,
superscript + for repetition more than once, superscript * for
repetition zero or more times, whereas using $\langle$...$\rangle$
with repetition denotes a comma separated list. Identifiers $v$, $p$
and $g$ denote variable names, patterns and guards, respectively, and
$e$ denotes an expression.
Note that $\{\}$ and $[]$ are parts of the syntax of Erlang representing tuples and lists, respectively.}
\label{ErlangSyntax}
\end{figure}



\begin{table}
\centering
\begin{tabular}{|rcl|}
  \hline
  \bf{Timed Rebeca} & & \bf{Erlang} \\
  \hline
  Model & $\rightarrow$ & A set of processes \\
  Reactive classes & $\rightarrow$ & A process whose behaviour consists of three functions \\
  Known rebecs & $\rightarrow$ & Record of variables \\
  State variables & $\rightarrow$ & Record of variables \\
  Message server & $\rightarrow$ & A match in a receive expression  \\
  Local variables & $\rightarrow$ & Record of variables \\
  Message send & $\rightarrow$ & Message send expression \\
  Message send w/after & $\rightarrow$ & Message send expression in the timeout block of a receive\\
    $~$ & $~$ &                          with an empty pattern, the timeout block is always executed, \\
  $~$ & $~$ &   sending the message after the specified time\\
    Message send w/deadline & $\rightarrow$ & Message send expression with the deadline as parameter \\
  Delay statement & $\rightarrow$ & Empty receive with a timeout \\
  Now expression & $\rightarrow$ & System time \\
  Assignment & $\rightarrow$ & Record update \\
  If statement & $\rightarrow$ & If expression \\
  Nondeterministic selection & $\rightarrow$ & Random selection in Erlang \\
  \hline
\end{tabular}
\caption{Structure of the mapping from Timed Rebeca to Erlang.}
\label{tab:erlmapping}
\end{table}

\begin{lstlisting}[float=tbp, caption=Pseudo Erlang code capturing the behaviour of the ticketService process., label=lst:procstate]
ticketService() ->
  receive
    % wait for a message with a set of known rebecs
    {Agent} ->
      % proceed to the next behaviour
      ticketService(#ticketService_knownrebecs{agent=Agent})
  end.
ticketService(KnownRebecs) ->
  receive
    % wait for the 'initial' message
    initial ->
      % process message 'initial' and proceed to the next behaviour
      ticketService(KnownRebecs, #ticketService_statevars{})
  end.
ticketService(KnownRebecs, StateVars) ->
  receive
    % wait for each message servers
    requestTicket ->
      % process message 'requestTicket' and loop
      ticketService(KnownRebecs, StateVars)
  end.
\end{lstlisting}

{Reactive classes} are translated into three functions, each representing
a possible behaviour of an Erlang process: 1)  the process waits to get references to known rebecs, 2) the process reads the initial message from the queue and executes it, 3) the process reads messages from the queue and executes them.
Once processes reach the last function they enter a loop. Erlang pseudocode
for the reactive class
\emph{TicketService} in the Rebeca model in Listing \ref{lst:timedticketservice} is shown in  Listing \ref{lst:procstate}.


A message server is translated into a match expression (see Figure
\ref{ErlangSyntax}), which is used inside \lstinline|receive ... end|. In
Listing \ref{lst:procstate}, \lstinline|requestTicket| is the pattern that is
matched on, and the body of the message server is mapped to the corresponding
expression.

Message send is implemented depending on whether \lstinline|after| is used. If
there is no \lstinline|after|, the message is sent like a regular message using
the \lstinline|!| operator, as shown on line 4 in Listing
\ref{lst:msgsendafter}. However, if the keyword \lstinline|after| is present a
new process is spawned which sleeps for the specified amount of time before
sending the message as described before.
Setting a deadline for the delivery of a message is possible by changing the
value \lstinline|inf|, which denotes no deadline (as shown on line 3 in Listing
\ref{lst:msgsendafter}), to an absolute point in time.
Messages are tagged with the time at which they were sent. For the simulation
we use the system clock to find out the current time by calling the Erlang
function \lstinline|now()|.

Moreover, since message servers can reply to the sender of the message, we need
to take care of setting the sender as part of the message as seen  on line 4 in Listing
\ref{lst:msgsendafter}.


\begin{lstlisting}[float=tbp, caption=Example of a message send after 15 time units in Erlang., label=lst:msgsendafter]
Sender = self(),
spawn(fun() ->
  receive after 15 ->
    TicketService ! {{Sender, now(), inf}, requestTicket}
  end
end)
\end{lstlisting}

As there is no pattern to match with, the \emph{delay} statement is
implemented as a receive consisting of just a timeout that makes the
process wait for a certain amount of time. For example,
\emph{delay}$(10)$ is translated to
\lstinline|receive after 10 ->  ok end|.

The \emph{deadline} of each message is checked right before the body
of the message server is executed. The current time is compared with
the deadline of the message to see if the deadline has expired and, if so,
the message is purged.
\section{Simulation of Timed Rebeca Using McErlang}

In this section, we present experimental results for two case studies.  The
first case study is the ticket service model displayed in Listing
\ref{lst:timedticketservice} and the second is a model of a sensor network.  In
each case we run a simulation for ten times, and for each case for 30 minutes or until a runtime monitor fails,
which means that an erroneous state has been reached. The simulations are run
in a setting in which a time unit is 1000 ms.
The experiment platform is Macbook 2.0GHz Intel Core 2 Duo - Aluminum 4GB memory Mac OS X, 10.6.6, and Erlang R13B04.


\paragraph{Ticket Service}
 The ticket service model is described in Section
\ref{sec:timedrebeca}. For each simulation,
we change one of the following parameters: the amount of time that is allowed to
pass before a request is processed, the time that passes before agent checks if he
has been issued a ticket, the amount of time that passes before agent tries the next
ticket service if he did not receive a ticket, the amount of time that passes before
agent restarts the ticket requests in case neither ticket service issued a ticket and
two different service times, which are non-deterministically chosen
as delay time in a ticket service and model the processing time for a request.
Table \ref{tab:experimentalresultsticketagency}  shows different settings of those parameters for which the ticket services never
issue a ticket to the agent because of tight deadlines, as well as settings for which a
ticket is issued during a simulation of the model.

\begin{table}[t]
\centering
\resizebox{\textwidth}{!}{
\begin{tabular}{|c c c c c c l|}
  \hline
  \rowH{Request}{deadline} & \rowH{Check issued}{period} & \rowH{Retry request}{period} & \rowH{New request}{period} & \rowH{Service}{time 1} & \rowH{Service}{time 2} & \rowh{Result} \\
  \hline
  2 & 1 & 1 & 1 & 3,4 & 7 & Not issued \\
  2 & 2 & 1 & 1 & 4 & 7 & Not issued \\
  2 & 2 & 1 & 1 & 3 & 7 & Ticket issued \\
  \hline
\end{tabular}
}
\label{tab:experimentalresultsticketagency}
\caption{Experimental simulation results for ticket service.}
\end{table}

\paragraph{Sensor Network}
We model a simple sensor network using Timed Rebeca. (See Listing
\ref{lst:sensornetwork} in Appendix \ref{AppendixB} for the complete
description of the model.)  A distributed sensor network is set up to
monitor levels of toxic gasses. The sensor rebecs (\verb|sensor0| and
\verb|sensor1|), announce the measured value to the admin node
(\verb|admin| rebec) in the network. If the admin node receives
reports of dangerous gas levels, it immediately notifies the scientist
(\verb|scientist| rebec) on the scene about it. If the scientist
does not acknowledge the notification within a given time frame, the
admin node sends a request to the rescue team (\verb|rescue| rebec) to
look for the scientist. The rescue team has a limited amount of time
units to reach the scientist and save him.


The rebecs \verb|sensor0| and \verb|sensor1| will periodically read
the gas-level measurement, modelled as a non-deterministic selection between
\verb|GAS_LOW| and \verb|GAS_HIGH|, and send their values to \verb|admin|.
The \verb|admin| continually checks, and acts
upon, the sensor values it has received.
When the \verb|admin| node receives a report of a reading that is life threatening for the
\verb|scientist| (\verb|GAS_HIGH|), it notifies him and waits for a limited amount of time units for an
acknowledgement.
The \verb|rescue| rebec represents a rescue team that is sent
off, should the \verb|scientist| not acknowledge the message from the
\verb|admin| in time. We model the response speed of the rescue team
with a non-deterministic delay of 0 or 1 time units.
%
The \verb|admin| keeps track
of the deadlines for the \verb|scientist| and the \verb|rescue| team
as follows:

\begin{itemize}
\item
  the \verb|scientist| must acknowledge that he is aware of a dangerous
  gas-level reading before {\em scientistDeadline} time units have passed;
\item
  the \verb|rescue| team must have reached the \verb|scientist| within
  {\em rescueDeadline} time units.
\end{itemize}
Otherwise we consider the mission failed.

The model can be parameterized over the values of network delay,
\verb|admin| sensor-read period, \verb|sensor0| read period,
\verb|sensor1| read period, scientist reply deadline and rescue-team
reply deadline, as shown in Table
\ref{tab:experimentalresultssensornetwork}.
\begin{table}[t]
\centering
\resizebox{\textwidth}{!}{
\begin{tabular}{|c c c c c c p{4cm}|}
  \hline
  \rowH{Network}{delay} & \rowH{Admin}{period} & \rowH{Sensor 0}{period} & \rowH{Sensor 1}{period} & \rowH{Scientist}{deadline} & \rowH{Rescue}{deadline} & \rowh{Result} \\
  \hline
  1       & 4       & 2       & 3       & 2       & 3       & Mission failed \\
  1       & 4       & 2       & 3       & 2       & 4       & Mission success \\
  2       & 1       & 1       & 1       & 4       & 5,6,7       & Mission failed \\
  2       & 4       & 1       & 1       & 4       & 7       & Mission success \\
  \hline
\end{tabular} 
}
\label{tab:experimentalresultssensornetwork}
\caption{Experimental simulation results for sensor network.}
\end{table}
In that table, we can see two
different cases in which we go from mission failure to mission success
between simulations. In the first scenario, we go from mission failure
to success as we increase the rescue deadline, as expected.
In the second scenario, we changed the parameters to model a faster sensor update and we observed mission failure. In this scenario,  increasing the rescue deadline further (from 5 to 7) is
insufficient. Upon closer inspection, we observe that our model fails
to cope with the rapid sensor updates and admin responses because it enters an unstable state.
The admin node initiates a new rescue mission while
another is still ongoing, eventually resulting in mission
failure. This reflects a design flaw in the model for frequent
updates that can be solved by keeping track of  an ongoing rescue mission in the model. Alternatively, increasing the value of \verb|admin| sensor-read period above half the rescue
deadline eliminates the flaw and the simulation is successful
again.

\section{Future Work}


The work reported in this paper paves the way to several interesting
avenues for future work.
In particular, we have already started modelling larger  real-world case studies and analyzing them using our tool.
We plan to explore different
approaches for model checking Timed Rebeca models. It is worth noting
that the translation from Timed Rebeca to Erlang immediately opens the
possibility of model checking untimed Rebeca models using
McErlang. This adds yet another component to the verification toolbox
for Rebeca, whose applicability needs to be analyzed via a series of
benchmark examples.
As mentioned in the paper, McErlang supports the notion of time only for simulation and not in model checking, and therefore cannot be used as is for model
checking Timed Rebeca models. We plan to explore different ways in
which McErlang can be used for model checking Timed Rebeca.
One possible solution is to store the local time of each process and write a custom-made scheduler in McErlang that simulates the way the Timed Rebeca scheduler operates.
The formal semantics for Timed Rebeca
presented in this paper is also used in another parallel line of work
\cite{javadMSThesis2010}. The aim of that study is to map Timed Rebeca
to timed automata \cite{Alur1994} in order to use UPPAAL
\cite{UppaalSite} for model checking Timed Rebeca models. The
translation from Timed Rebeca to timed automata will be integrated in
our tool suite. 
We are also working on a translation of Timed Rebeca into (Real-time) Maude. This alternative translation would allow designers to use the analysis tools supported by Maude in the verification and validation of Timed Rebeca models.
Our long-term goal is to have a tool suite for
modelling, executing, simulating, and model checking asynchronous
object-based systems using Timed Rebeca.

\paragraph{Acknowledgements}
{The work on this paper has been partially supported by the projects
``New Developments in Operational Semantics'' (nr.~080039021),
``Meta-theory of Algebraic Process Theories'' (nr.~100014021) and
``Timed Asynchronous Reactive Objects in Distributed Systems: TARO''
(nr.~110020021) of the Icelandic Research Fund.}

\bibliographystyle{eptcs}

\begin{thebibliography}{10}
\providecommand{\bibitemdeclare}[2]{}
\providecommand{\urlprefix}{Available at }
\providecommand{\url}[1]{\texttt{#1}}
\providecommand{\href}[2]{\texttt{#2}}
\providecommand{\urlalt}[2]{\href{#1}{#2}}
\providecommand{\doi}[1]{doi:\urlalt{http://dx.doi.org/#1}{#1}}
\providecommand{\bibinfo}[2]{#2}

\bibitemdeclare{book}{Agha86}
\bibitem{Agha86}
\bibinfo{author}{G.~Agha} (\bibinfo{year}{1990}):
  \emph{\bibinfo{title}{{Actors}: {A} Model of Concurrent Computation in
  Distributed Systems}}.
\newblock \bibinfo{publisher}{MIT Press, Cambridge, MA, USA}.

\bibitemdeclare{article}{Alur1994}
\bibitem{Alur1994}
\bibinfo{author}{R.~Alur} \& \bibinfo{author}{D~Dill} (\bibinfo{year}{1994}):
  \emph{\bibinfo{title}{A Theory of Timed Automata}}.
\newblock {\sl \bibinfo{journal}{Theoretical Computer Science}}
  \bibinfo{volume}{126}, pp. \bibinfo{pages}{183--235}.
\newblock \doi{10.1016/0304-3975(94)90010-8}.

\bibitemdeclare{techreport}{Baker78actorsystems}
\bibitem{Baker78actorsystems}
\bibinfo{author}{Henry~Givens Baker} (\bibinfo{year}{1978}):
  \emph{\bibinfo{title}{Actor Systems for Real-Time Computation}}.
\newblock \bibinfo{type}{Technical Report}, \bibinfo{institution}{MIT}.

\bibitemdeclare{inproceedings}{DBLP:journals/corr/abs-1009-4262}
\bibitem{DBLP:journals/corr/abs-1009-4262}
\bibinfo{author}{Joakim Bj{\o}rk}, \bibinfo{author}{Einar~Broch Johnsen},
  \bibinfo{author}{Olaf Owe} \& \bibinfo{author}{Rudolf Schlatte}
  (\bibinfo{year}{2010}): \emph{\bibinfo{title}{Lightweight Time Modeling in
  {Timed Creol}}}.
\newblock In: {\sl \bibinfo{booktitle}{RTRTS}}, pp. \bibinfo{pages}{67--81}.
\newblock \doi{10.4204/EPTCS.36.4}.

\bibitemdeclare{inproceedings}{DBLP:conf/fsen/BoerCJ09}
\bibitem{DBLP:conf/fsen/BoerCJ09}
\bibinfo{author}{Frank~S. de~Boer}, \bibinfo{author}{Tom Chothia} \&
  \bibinfo{author}{Mohammad~Mahdi Jaghoori} (\bibinfo{year}{2009}):
  \emph{\bibinfo{title}{Modular Schedulability Analysis of Concurrent Objects
  in {Creol}}}.
\newblock In: {\sl \bibinfo{booktitle}{FSEN}}, pp. \bibinfo{pages}{212--227}.
\newblock \doi{10.1007/978-3-642-11623-0\_12}.

\bibitemdeclare{misc}{ErlangSite}
\bibitem{ErlangSite}
\bibinfo{author}{{Erlang}}: \emph{\bibinfo{title}{{Erlang Programming Language
  Homepage}}}.
\newblock \bibinfo{note}{Http://www.erlang.org}.

\bibitemdeclare{inproceedings}{McErlang}
\bibitem{McErlang}
\bibinfo{author}{Lars-{\AA}ke Fredlund} \& \bibinfo{author}{Hans Svensson}
  (\bibinfo{year}{2007}): \emph{\bibinfo{title}{McErlang: a model checker for
  a distributed functional programming language}}.
\newblock In: {\sl \bibinfo{booktitle}{ICFP}}, pp. \bibinfo{pages}{125-136}.
\newblock \doi{10.1145/1291151.1291171}.

\bibitemdeclare{techreport}{Hewitt72}
\bibitem{Hewitt72}
\bibinfo{author}{C.~Hewitt} (\bibinfo{year}{1972}):
  \emph{\bibinfo{title}{Description and Theoretical Analysis (Using Schemata)
  of {PLANNER}: {A} Language for Proving Theorems and Manipulating Models in a
  Robot}}.
\newblock \bibinfo{type}{{MIT} Artificial Intelligence Technical Report}
  \bibinfo{number}{258}, \bibinfo{institution}{Department of Computer Science,
  {MIT}}.

\bibitemdeclare{inproceedings}{Hewitt07-commitment}
\bibitem{Hewitt07-commitment}
\bibinfo{author}{Carl Hewitt} (\bibinfo{year}{2007}):
  \emph{\bibinfo{title}{What is Commitment? {Physical}, Organizational, and
  Social (Revised)}}.
\newblock In: {\sl \bibinfo{booktitle}{Proceedings of Coordination,
  Organizations, Institutions, and Norms in Agent Systems II}},
  \bibinfo{series}{Lecture Notes in Computer Science},
  \bibinfo{publisher}{Springer}, pp. \bibinfo{pages}{293--307}.
\newblock \doi{10.1007/978-3-540-74459-7\_19}.

\bibitemdeclare{misc}{IceRoseSite}
\bibitem{IceRoseSite}
\bibinfo{author}{{ICEROSE}}: \emph{\bibinfo{title}{{ICEROSE Homepage}}}.
\newblock
  \bibinfo{note}{Http://en.ru.is/icerose/applying-formal-methods/projects/TARO}.

\bibitemdeclare{misc}{javadMSThesis2010}
\bibitem{javadMSThesis2010}
\bibinfo{author}{Mohammad~Javad Izadi} (\bibinfo{year}{2010}):
  \emph{\bibinfo{title}{An Actor-based Model for Modeling and Verification of
  Real-Time Systems - {Master Thesis, University of Tehran, Iran}}}.

\bibitemdeclare{inproceedings}{Jaghoori07nwpt}
\bibitem{Jaghoori07nwpt}
\bibinfo{author}{M.~M. Jaghoori}, \bibinfo{author}{F.S. de~Boer},
  \bibinfo{author}{T.~Chothia} \& \bibinfo{author}{M.~Sirjani}
  (\bibinfo{year}{2007}): \emph{\bibinfo{title}{Task scheduling in {Rebeca}}}.
\newblock In: {\sl \bibinfo{booktitle}{Proc. Nordic Workshop on Programming
  Theory (NWPT'07)}}.
\newblock \bibinfo{note}{Extended abstract}.

\bibitemdeclare{article}{Jaghoori08jlap}
\bibitem{Jaghoori08jlap}
\bibinfo{author}{M.~M. Jaghoori}, \bibinfo{author}{F.S. de~Boer},
  \bibinfo{author}{T.~Chothia} \& \bibinfo{author}{M.~Sirjani}
  (\bibinfo{year}{2009}): \emph{\bibinfo{title}{Schedulability of Asynchronous
  Real-Time Concurrent Objects}}.
\newblock {\sl \bibinfo{journal}{Logic and Algebraic Programming}}
  \bibinfo{volume}{78}(\bibinfo{number}{5}), pp. \bibinfo{pages}{402--416}.
\newblock \bibinfo{note}{A preliminary version appeared in NWPT/FLACOS 2007 as
  an extended abstract}.
\newblock \doi{10.1016/j.jlap.2009.02.009}.

\bibitemdeclare{article}{jaghoor-sirjani-mousavi-movaghar-09-Acta}
\bibitem{jaghoor-sirjani-mousavi-movaghar-09-Acta}
\bibinfo{author}{Mohammad~Mahdi Jaghoori}, \bibinfo{author}{Marjan Sirjani},
  \bibinfo{author}{Mohammad~Reza Mousavi}, \bibinfo{author}{Ehsan Khamespanah}
  \& \bibinfo{author}{Ali Movaghar} (\bibinfo{year}{2009}):
  \emph{\bibinfo{title}{Symmetry and Partial Order Reduction Techniques in
  Model Checking {Rebeca}}}.
\newblock {\sl \bibinfo{journal}{Acta Informaticae}}
  \bibinfo{volume}{47}(\bibinfo{number}{1}), pp. \bibinfo{pages}{33--66}.
\newblock \doi{10.1007/s00236-009-0111-x}.

\bibitemdeclare{inproceedings}{Kahn87}
\bibitem{Kahn87}
\bibinfo{author}{Gilles Kahn} (\bibinfo{year}{1987}):
  \emph{\bibinfo{title}{Natural Semantics}}.
\newblock In \bibinfo{editor}{Franz-Josef Brandenburg}, \bibinfo{editor}{Guy
  Vidal-Naquet} \& \bibinfo{editor}{Martin Wirsing}, editors: {\sl
  \bibinfo{booktitle}{STACS 87, 4th Annual Symposium on Theoretical Aspects of
  Computer Science, Passau, Germany, February 19-21, 1987, Proceedings}}, {\sl
  \bibinfo{series}{Lecture Notes in Computer Science}} \bibinfo{volume}{247},
  \bibinfo{publisher}{Springer-Verlag}, pp. \bibinfo{pages}{22--39}.
\newblock \doi{10.1007/BFb0039592}.

\bibitemdeclare{inproceedings}{Nielsen96}
\bibitem{Nielsen96}
\bibinfo{author}{Brian Nielsen} \& \bibinfo{author}{Gul Agha}:
  \emph{\bibinfo{title}{Semantics for an actor-based real-time language}}.
\newblock In: {\sl \bibinfo{booktitle}{Proceedings of The Fourth International
  Workshop on Parallel and Distributed Real-Time Systems (WPDRS'96)}},
  \bibinfo{publisher}{IEEE Computer Society Press, Los Alamitos, CA, USA,
  1996}.

\bibitemdeclare{inproceedings}{NigroP01}
\bibitem{NigroP01}
\bibinfo{author}{Libero Nigro} \& \bibinfo{author}{Francesco Pupo}
  (\bibinfo{year}{2001}): \emph{\bibinfo{title}{Schedulability Analysis of Real
  Time Actor Systems Using Coloured Petri Nets}}.
\newblock In: {\sl \bibinfo{booktitle}{Proc. Concurrent Object-Oriented
  Programming and Petri Nets}}, pp. \bibinfo{pages}{493--513}.
\newblock \doi{10.1007/3-540-45397-0\_21}.

\bibitemdeclare{article}{OlveczkyM02}
\bibitem{OlveczkyM02}
\bibinfo{author}{Peter~Csaba {\"O}lveczky} \& \bibinfo{author}{Jos{\'e}
  Meseguer} (\bibinfo{year}{2002}): \emph{\bibinfo{title}{Specification of
  real-time and hybrid systems in rewriting logic}}.
\newblock {\sl \bibinfo{journal}{Theor. Comput. Sci.}}
  \bibinfo{volume}{285}(\bibinfo{number}{2}), pp. \bibinfo{pages}{359--405}.
\newblock \doi{10.1016/S0304-3975(01)00363-2}.

\bibitemdeclare{techreport}{Plotkin81}
\bibitem{Plotkin81}
\bibinfo{author}{G.~D. Plotkin} (\bibinfo{year}{1981}): \emph{\bibinfo{title}{A
  Structural Approach to Operational Semantics}}.
\newblock \bibinfo{type}{Technical Report} \bibinfo{number}{DAIMI FN-19},
  \bibinfo{institution}{Computer Science Department, Aarhus University},
  \bibinfo{address}{Aarhus, Denmark}.

\bibitemdeclare{inproceedings}{RenA95}
\bibitem{RenA95}
\bibinfo{author}{Shangping Ren} \& \bibinfo{author}{Gul Agha}
  (\bibinfo{year}{1995}): \emph{\bibinfo{title}{{RT-synchronizer}: Language
  Support for Real-Time Specifications in Distributed Systems}}.
\newblock In: {\sl \bibinfo{booktitle}{Workshop on Languages, Compilers and
  Tools for Real-Time Systems}}, pp. \bibinfo{pages}{50--59}.
\newblock \doi{10.1145/216636.216656}.

\bibitemdeclare{article}{Sirjani05JUCS}
\bibitem{Sirjani05JUCS}
\bibinfo{author}{M.~Sirjani}, \bibinfo{author}{A.~Movaghar},
  \bibinfo{author}{A.~Shali} \& \bibinfo{author}{F.S. de~Boer}
  (\bibinfo{year}{2005}): \emph{\bibinfo{title}{Model Checking, Automated
  Abstraction, and Compositional Verification of {Rebeca} Models}}.
\newblock {\sl \bibinfo{journal}{Journal of Universal Computer Science}}
  \bibinfo{volume}{11}(\bibinfo{number}{6}), pp. \bibinfo{pages}{1054--1082}.

\bibitemdeclare{article}{Sirjani04FI}
\bibitem{Sirjani04FI}
\bibinfo{author}{M.~Sirjani}, \bibinfo{author}{A.~Movaghar},
  \bibinfo{author}{A.~Shali} \& \bibinfo{author}{F.S. de~Boer}
  (\bibinfo{year}{Dec. 2004}): \emph{\bibinfo{title}{Modeling and Verification
  of Reactive Systems using {Rebeca}}}.
\newblock {\sl \bibinfo{journal}{Fundamenta Informatica}}
  \bibinfo{volume}{63}(\bibinfo{number}{4}), pp. \bibinfo{pages}{385--410}.

\bibitemdeclare{book}{Tanenbaum2002}
\bibitem{Tanenbaum2002}
\bibinfo{author}{Andrew~S. Tanenbaum} \& \bibinfo{author}{Maarten van Steen}
  (\bibinfo{year}{2007}): \emph{\bibinfo{title}{Distributed systems -
  principles and paradigms (2. ed.)}}.
\newblock \bibinfo{publisher}{Pearson Education}.

\bibitemdeclare{misc}{UppaalSite}
\bibitem{UppaalSite}
\bibinfo{author}{{UPPAAL}}: \emph{\bibinfo{title}{{UPPAAL Homepage}}}.
\newblock \bibinfo{note}{Http://uppaal.com}.

\bibitemdeclare{inproceedings}{YiTCCS91b}
\bibitem{YiTCCS91b}
\bibinfo{author}{Wang Yi} (\bibinfo{year}{1991}): \emph{\bibinfo{title}{{CCS} +
  time = an interleaved model for real time systems}}.
\newblock In: {\sl \bibinfo{booktitle}{Proceedings of ICALP 1991}}, {\sl
  \bibinfo{series}{Lecture Notes in Computer Science}} \bibinfo{volume}{510},
  \bibinfo{publisher}{Springer-Verlag}, pp. \bibinfo{pages}{217--228}.
\newblock \doi{10.1007/3-540-54233-7\_136}.

\end{thebibliography}

\newpage
\appendix

\section{Method-Execution Transition Rules}\label{App:methodrules}   

\begin{figure}[h]
\begin{align*}
(msg) & ~ \axiomrule{(varname.m(\overline{v}) ~ after(d) ~deadline( DL), \sigma, Env, B)} \\
& ~ \axiomrule{\trans{\tau} (\sigma, Env, \{(\sigma(varname), m(eval(\overline{v}, \sigma)), \sigma(self), \sigma(now)+d, \sigma(now)+DL)\} \cup B)} \\
\\
(delay)& ~ \axiomrule{(delay(d), \sigma, Env, B) \trans{\tau} (\sigma[now = \sigma(now) + d], Env, B)} \\
\\
(assign) & ~ \axiomrule{(x=e, \sigma, Env, B) \trans{\tau} (\sigma[x=eval(e,\sigma)], Env, B)}\\
\\
(create)& ~ \axiomrule{(varname = new~ O(\overline{v}), \sigma, Env, B)} \\
& ~ \axiomrule{\trans{\tau} (\sigma[varname = A],  \{\sigma_{A}[now = \sigma(now), ~ \mathit{self} = A]\} \cup Env, \{(A, initial(eval(\overline{v}, \sigma)), \sigma(self)),\sigma(now), +\infty)\} \cup B)} \\
\\
(cond_1)& ~ \sosrule{eval(e, \sigma) = true \quad (S_1, \sigma, Env,  B) \trans{\tau} (\sigma', Env',  B')}{(if ~(e)~ then~ S_1~ else~ S_2, \sigma, Env, B) \trans{\tau} (\sigma', Env', B')}\\
\\
(cond_2)& ~\sosrule{eval(e, \sigma) = false \quad (S_2, \sigma, Env,  B) \trans{\tau} (\sigma', Env',  B')}{(if ~(e)~ then~ S_1~ else~ S_2,~ \sigma,Env, B) \trans{\tau} (\sigma', Env', B')}\\
\\
(seq)& ~\sosrule{(S_1, \sigma, Env,  B) \trans{\tau} (\sigma', Env',  B'), ~(S_2, \sigma', Env', B') \trans{\tau} (\sigma'', Env'',  B'')}{(S_1;S_2, \sigma, Env, B) \trans{\tau} (\sigma'', Env'', B'')}
\end{align*}
\caption{The Method-Execution Transitions Rules. In rule {\em create}, the rebec name $A$ should not appear in the range of the environment $\sigma$. The function $eval$ evaluates expressions in a given environment in the expected way. In each rule, we assume that $\sigma$ is not contained in $Env$.}
\label{fig:fulltransitions}
\end{figure}

\section{Rebeca Model for the Sensor Network} \label{AppendixB}

\lstset{language=rebeca,style=scrsize}
\begin{lstlisting}[language=rebeca,caption=A Timed Rebeca model of the sensor network example,label=lst:sensornetwork]
env int netDelay;
env int adminCheckDelay;
env int sensor0period;
env int sensor1period;
env int scientistDeadline;
env int rescueDeadline;

reactiveclass Sensor {
    knownrebecs {
        Admin admin;
    }

    statevars {
        int period;
    }
    
    msgsrv initial(int myPeriod) {
        period = myPeriod;
        self.doReport();
    }
    
    msgsrv doReport() {
        int value;
        value = ?(2, 4); // 2=safe gas levels, 4=danger gas levels
        admin.report(value) after(netDelay);
        self.doReport() after(period);
    }
}

reactiveclass Scientist {
    knownrebecs {
        Admin admin;
    }

    msgsrv initial() {}
    
    msgsrv abortPlan() {
        admin.ack() after(netDelay);
    }
}

reactiveclass Rescue {
    knownrebecs {
        Admin admin;
    }

    msgsrv initial() {}
    
    msgsrv go() {
        int msgDeadline = now() + (rescueDeadline-netDelay);
        int excessiveDelay = ?(0, 1); // unexpected obstacle might occur during rescue
        delay(excessiveDelay);
        admin.rescueReach() after(netDelay) deadline(msgDeadline);
    }
}

reactiveclass Admin {
    knownrebecs {
        Sensor sensor0;
        Sensor sensor1;
        Scientist scientist;
        Rescue rescue;
    }

    statevars {
        boolean reported0;
        boolean reported1;
        int sensorValue0;
        int sensorValue1;
        boolean sensorFailure;
        boolean scientistAck;
        boolean scientistReached;
        boolean scientistDead;
    }
    
    msgsrv initial() {
        self.checkSensors();
    }
    
    msgsrv report(int value) {
        if (sender == sensor0) {
            reported0 = true;
            sensorValue0 = value;
        } else {
            reported1 = true;
            sensorValue1 = value;
        }
    }
    
    msgsrv rescueReach() {
        scientistReached = true;
    }
    
    msgsrv checkSensors() {
        if (reported0) reported0 = false;
        else sensorFailure = true;

        if (reported1) reported1 = false;
        else sensorFailure = true;

        boolean danger = false;
        if (sensorValue0 > 3) danger = true;
        if (sensorValue1 > 3) danger = true;

        if (danger) {
            scientist.abortPlan() after(netDelay);
            self.checkScientistAck() after(scientistDeadline); // deadline for the scientist to answer
        }

        self.checkSensors() after(adminCheckDelay);
    }
    
    msgsrv checkRescue() {
        if (!scientistReached) {
            scientistDead = true; // scientist is dead
        } else {
            scientistReached = false;
        }
    }
    
    msgsrv ack() {
        scientistAck = true;
    }
    
    msgsrv checkScientistAck() {
        if (!scientistAck) {
            rescue.go() after(netDelay);
            self.checkRescue() after(rescueDeadline);
        }
        scientistAck = false;
    }
}

main {
    Sensor sensor0(admin):(sensor0period);
    Sensor sensor1(admin):(sensor1period);
    Scientist scientist(admin):();
    Rescue rescue(admin):();
    Admin admin(sensor0, sensor1, scientist, rescue):();
}

\end{lstlisting}


\end{document}